# Symmetry breaking of the bending mode of CO$_2$ in the presence of Ar

T.A. Gartner,[1] A.J. Barclay,[1] A.R.W. McKellar,[2] and N. Moazzen-Ahmadi[1]

[1] *Department of Physics and Astronomy, University of Calgary, 2500 University Drive North West, Calgary, Alberta T2N 1N4, Canada*

[2] *National Research Council of Canada, Ottawa, Ontario K1A 0R6, Canada*

**Abstract**

The weak infrared spectrum of CO$_2$-Ar corresponding to the $(01^11) \leftarrow (01^10)$ hot band of CO$_2$ is detected in the region of the carbon dioxide $\nu_3$ fundamental vibration ($\approx 2340$ cm$^{-1}$), using a tunable OPO laser source to probe a pulsed supersonic slit jet expansion. While this method was previously thought to cool clusters to the lowest rotational states of the ground vibrational state, here we show that under suitable jet expansion conditions, sufficient population remains in the first excited bending mode of CO$_2$ (1-2%) to enable observation of vibrationally hot CO$_2$-Ar, and thus to investigate the symmetry breaking of the intramolecular bending mode of CO$_2$ in the presence of Ar. The bending mode of CO$_2$ monomer splits into an in-plane and an out-of-plane mode, strongly linked by a Coriolis interaction. Analysis of the spectrum yields a direct measurement of the in-plane / out-of-plane splitting measured to be 0.8770 cm$^{-1}$. Calculations were carried to determine if key features of our results, i.e., the sign and magnitude of the shift in the energy for the two intramolecular bending modes, are consistent with a quantum chemical potential energy surface. This aspect of intramolecular interactions has received little previous experimental and theoretical consideration. Therefore, we provide an additional avenue by which to study the intramolecular dynamics of this simplest dimer in its bending modes. Similar results should be possible for other weakly-bound complexes.

**1. Introduction**

High resolution spectroscopy of weakly-bound van der Waals complexes gives direct information on intermolecular forces and dynamics. Resolved rotational transitions give structural information which helps to determine minima in the intermolecular potentials. Intermolecular vibrational intervals give complementary information on the potential energy surface. Intramolecular vibrations exhibit shifts (usually relatively small) which show how the monomer dynamics are affected by the presence of a weakly-bound partner. Most current information on such vibrational shifts relates to stretching vibrations; there is very little data for monomer bends, in particular for the bending modes of simple linear molecules like $CO_2$, $N_2O$, $C_2H_2$, HCN, etc. This shortage of data is probably due in part to a lack of good tunable laser sources in the far-infrared region (say below 1000 cm$^{-1}$) where these bending fundamentals usually lie. One notable exception to this data shortage is a 1993 study of the $C_2H_2$-Ar dimer by Ohshima et al.[1] They studied the spectrum in the region of the $C_2H_2$ $\nu_5$ bending vibration ($\approx$730 cm$^{-1}$), and found that the presence of the nearby argon atom induces an overall blue shift of +0.393 cm$^{-1}$ and a modest splitting of 0.138 cm$^{-1}$.

In the present paper, we study the $\nu_2$ bending mode of $CO_2$, $(v_1, v_2^{l_2}, v_3) = (01^10)$, in the $CO_2$-Ar dimer. Rather than directly observing the $CO_2$ $\nu_2$ fundamental in the 667 cm$^{-1}$ region, we detect the $(01^11) \leftarrow (01^10)$ hot band in the $\nu_3$ fundamental region near 2337 cm$^{-1}$. This serendipitous observation is due to the inefficiency in vibrational cooling in the supersonic jet which leaves 1-2% of the vibrationally hot $CO_2$-Ar in the first excited bending mode of $CO_2$. We therefore face the difficulty of a small population in our $(01^10)$ initial state. But this is compensated by the advantages of an excellent tunable mid-infrared source, the use of quantum-correlated twin beams for detection[2] and a large transition strength for the $\nu_3$ mode.

Spectra of $CO_2$-Ar were first studied in the microwave region by Steed et al.,[3] who established a T-shaped structure with an intermolecular separation of about 3.5 Å. Since then, there has been further microwave work,[4,5] as well as infrared studies in the $CO_2$ $\nu_3$ [6-10] and $\nu_3 + \nu_1/2\nu_2$ [4,11] regions. The $CO_2$-$Ar_2$ trimer has also been detected.[12,13] There have been many theoretical investigations of $CO_2$-Ar interactions,[11-19] including two relatively recent *ab initio* studies of the complete potential surface which explicitly include its dependence on the $CO_2$ $\nu_3$ mode.[20,21] But as far as we know, there are no theoretical predictions of the shift or splitting of the $CO_2$ $\nu_2$ mode in $CO_2$-Ar.

## 2. Background theory

The $CO_2$-Ar structure is T-shaped, with the *a*-inertial axis connecting Ar and C, the *b*-axis parallel to the O-C-O axis, and the *c*-axis perpendicular to the $CO_2$-Ar plane. This corresponds to a $C_{2v}$ point group, whose fundamental vibrational modes can have $A_1$ symmetry (intramolecular $\nu_1$ symmetric stretch, in-plane component of the intramolecular $\nu_2$ bend, intermolecular van der Waals stretch), $B_2$ symmetry (intramolecular $\nu_3$ stretch, intermolecular bend), or $B_1$ symmetry (out-of-plane component of the intramolecular $\nu_2$ bend). In addition, $A_2$ symmetry can occur for combination modes involving $B_1 \times B_2$. Nuclear spin statistics for $^{16}O$ mean that only even-$K_a$ levels are possible for $A_1$ and $A_2$ modes (including the ground state), and only odd-$K_a$ for $B_1$ and $B_2$ modes. In this paper, $K$ stands for $K_a$ unless otherwise indicated.

In the present case, we are interested in the in-plane (i-p) and out-of-plane (o-p) $CO_2$-Ar vibrations corresponding to the $CO_2$ $(01^10)$ and $(01^11)$ modes. As already mentioned, the lower states, the $(01^10)$ i-p and o-p modes, have $A_1$ and $B_1$ symmetry, respectively. For the upper states, the $(01^11)$ i-p and o-p modes, we multiply by $B_2$ and obtain $B_2$ and $A_2$ symmetry, respectively.

The $CO_2$ $\nu_3$ stretch generates a transition moment parallel to the $CO_2$-Ar $b$-axis, so we expect spectra with $b$-type selection rules ($\Delta K_a = \pm 1$, $\Delta K_c = \pm 1, \pm 3$). The i-p component will be $B_2 \leftarrow A_1$ with only $K_a$ = odd ← even subbands, and the o-p component will be $A_2 \leftarrow B_1$ with only $K_a$ = even ← odd subbands. We also expect strong Coriolis mixing between the i-p and o-p modes within the lower and upper states, just as observed by Ohshima et al.[1] for $C_2H_2$-Ar. In our case this is a $b$-type interaction, characterized by a matrix element

$$\langle \text{i-p}, J, k \mid H \mid \text{o-p}, J, k \pm 1 \rangle = \tfrac{1}{2}\, \xi_b \times [J(J+1) - k(k \pm 1)]^{1/2},$$

where $k$ is signed $K_a$, and $\xi_b$ is the Coriolis interaction parameter (notation used for consistency with Ref. 1). The parameter $\xi_b$ is connected to the usual dimensionless Coriolis zeta parameter by the relation $\xi_b = 2B\zeta$, where $B$ is the $B$ rotational constant and $\zeta$ can take values between zero (no coupling) and unity (complete Coriolis coupling).

### 3. Results

#### 3.1. Experimental spectrum analysis

Spectra were recorded at the University of Calgary as described previously,[22-24] using a pulsed supersonic slit jet expansion probed by a rapid-scan optical parametric oscillator source and the use of quantum-correlated twin beams (idler and signal) for cancellation of the power fluctuations.[2] The gas expansion mixture contained about 0.04% carbon dioxide plus 0.8% argon in helium carrier gas with a backing pressure of about 13 atmospheres. Spectra were also recorded with a mixture containing only $CO_2$ and He to help verify which spectral features required Ar. Wavenumber calibration was carried out by simultaneously recording signals from a fixed etalon and a reference gas cell containing room temperature $CO_2$. Spectral simulation and fitting were made using the PGOPHER software.[25]

The observed spectrum is shown in the top trace of Fig. 1, where known transitions of $CO_2$ monomer have been 'clipped out' in order to show only features requiring the presence of Ar. The central $K = 1 \leftarrow 0$ subband of the i-p component is quite prominent in Fig. 1, with its strong $Q$-branch (2336.35 - 2336.50 cm$^{-1}$) and $R$-branch (2336.6 - 2337.3 cm$^{-1}$). But the correct assignment of this subband (and the rest of the spectrum) was not immediately evident. Our assignment strategy was to fix all rotational constants at their known ground state values,[3,6] and to assume complete Coriolis mixing in both the lower and upper states (that is, assume $\zeta = 1$, meaning that $\xi_b = 2B = 0.132$ cm$^{-1}$). This reduced the search for a good PGOPHER simulation to one unknown parameter, namely the separation of the i-p and o-p modes, assumed to be equal in the lower and upper states. When the correct solution was found, it turned out that the $K = 1 \leftarrow 0$ $Q$-branch is shaded such that higher $J$-values go to lower wavenumbers, opposite to the normal expectation for a $b$-type band. This effect of the Coriolis interaction helps to explain why the correct assignment was not obvious at first.

Ultimately, we fitted 178 transitions with values of $J$ ranging up to 11, and values of $K$ from 0 to 3 in both the upper and lower states. The resulting parameters are shown in Table 1. This Table includes known[6] ground state $CO_2$-Ar parameters for comparison. The quality of the fit is very good, with an average rms deviation (obs – calc) of about 0.0002 cm$^{-1}$, similar to the experimental uncertainty. In addition to the normally expected transitions within the i-p or o-p stacks of rotational levels, we also observed many transitions between the stacks. Such transitions are made possible by the Coriolis mixing, and follow selection rules $\Delta K_a = 0, \pm 2$, $\Delta K_c = 0, \pm 2$. They of course ensure that the splittings between i-p and o-p level stacks are very accurately determined. The division of our simulated spectra (Fig. 1) into in-plane and out-of-plane components, as given by the PGOPHER fit, is somewhat arbitrary because the Coriolis mixing is so strong.

### 3.2. Discussion

In order to illustrate the Coriolis interaction in $CO_2$-Ar, Fig. 2 shows lower and upper state calculated energy levels with and without the interaction. These are "reduced" energy levels from which the basic rotational dependence, $½(B + C) \times J(J + 1)$, has been subtracted for clearer presentation. An important point is that, because the o-p mode is shifted up relative to the i-p mode, the o-p $K = 1$ levels in the lower state lie very close to the i-p $K = 2$ levels. This greatly enhances the effect of the Coriolis interaction, leading to almost complete mixing of $K = 1$ and $2$ as $J$ increases. Levels with $K = 0$ encounter smaller, but still significant shifts, while $K = 3$ levels are only slightly perturbed. The level pattern is quite different in the $CO_2$ ($01^11$) upper vibrational state, even though its parameters such as splitting are very similar, since now the i-p mode has odd $K$-values and the o-p mode has even $K$.

For the bending mode of an isolated linear molecule such as $CO_2$, it is most useful to characterize vibrational dynamics in terms of vibrational angular momentum, equivalent to a Coriolis interaction with an implicit $\zeta$-value of exactly unity. Since the differential perturbation introduced by Ar in the $CO_2$ $\nu_2$ vibration is very small (namely 0.88 in 667 cm$^{-1}$), or in other words, the degeneracy of the bending vibration is only slightly lifted. Thus it is not surprising that we observe a $\zeta$-value close to 1.0 for $CO_2$-Ar. In fact, the $\xi_b$ parameters from Table 1 result in values of about 1.01 for $\zeta$, depending slightly on exactly which value is used for $B$. A value greater than unity is not normally possible within the harmonic approximation, but we can ascribe the slightly larger value found here to the floppy nature of $CO_2$-Ar. In the case of $C_2H_2$-Ar, Ohshima et al.[1] determined a more anomalous $\zeta$-value of about 1.52, and also found that their fit required relatively large second order and centrifugal distortion corrections to $\xi$. Such corrections were not required here. These differences between $CO_2$-Ar and $C_2H_2$-Ar are consistent with the fact that $C_2H_2$-Ar is considerably less rigid, as manifested ~~shown~~ for example by its much lower energy intermolecular bending mode (6.2 cm$^{-1}$)[26] compared to $CO_2$-Ar (27.8 cm$^{-1}$).[8]

We estimate the strongest lines in the observed $CO_2$-Ar $(01^11) \leftarrow (01^10)$ hot band to be about 0.022 times the strength of those in the $(001) \leftarrow (000)$ fundamental band, though this is rather uncertain due to possible changes in laser output power and supersonic jet conditions. The jet nozzle is at room temperature, where the population in each of the degenerate $(01^10)$ modes of $CO_2$ is about 0.038 times that of the $(000)$ ground state, for a total of 0.076. Following the supersonic expansion and formation of $CO_2$-Ar, the effective rotational temperature is about 1.8 K, and some of the population of the o-p mode has relaxed into the lower energy i-p mode, so that the i-p component of the spectrum is stronger, as seen in Fig. 1. Putting these observations together, we can say that the observed strength of the $CO_2$-Ar $(01^11) \leftarrow (01^10)$ hot band is about half (very roughly!) of the maximum value that would be observed if all the original $(01^10)$ state population of $CO_2$ survived the supersonic expansion without any relaxation. Thus it seems that the vibrational relaxation of the $(01^10)$ mode is relatively inefficient. This is probably due to the fact that our expansion gas mixture is mostly (>99%) helium, and that He – $CO_2$ collisions are relatively inefficient at removing the 667 cm$^{-1}$ energy of the $CO_2$ $\nu_2$ vibration.

The key parameter in our results is the splitting between i-p and o-p modes, which is determined to be 0.8773(1) cm$^{-1}$ in the lower $(01^10)$ state and 0.8764(1) cm$^{-1}$ in the upper $(01^11)$ state, with i-p lying below o-p. This rather basic aspect of intermolecular interactions in weakly-bound complexes containing linear molecules has received little or no theoretical consideration in the past, especially considering the many theoretical studies of $CO_2$-Ar. The only similar experimental result,[1] for $\nu_5$ of $C_2H_2$-Ar, involved the same ordering (i-p below o-p), but with a smaller splitting of just 0.138 cm$^{-1}$. Because of the nature of our spectrum, we only determine the splittings, not the overall shift of the $CO_2$ $\nu_2$ mode in $CO_2$-Ar (this overall shift is contained in the unknown parameter X in Table 1). But we do determine the shift of the $(01^11) \leftarrow (01^10)$ hot band origin relative to the free $CO_2$ monomer,[27] which is -0.464 cm$^{-1}$. This shift is virtually the same for the i-p and o-p modes, and is very similar to that of the $(001) \leftarrow (000)$ fundamental band, -0.470 cm$^{-1}$, as determined by Randall et al.[6] This similarity and the similarity of the splitting in the lower and upper states indicate that the coupling of the $\nu_2$ and $\nu_3$ modes in $CO_2$ remains almost unaffected in $CO_2$-Ar.

How can we understand the sign and magnitude of the symmetry breaking splitting? Compared to the o-p mode, the i-p mode seemingly involves moving the C atom to "crash into" the Ar atom once every vibration. So, one might think that the relatively steep repulsive wall of the intermolecular potential would tend to increase the i-p frequency relative to the o-p, contrary to our observation. We undertook some straightforward *ab initio* calculations to investigate this question. Using Gaussian 16 at the MP2/cc-pvtz level, we obtained values of 654.8 and 655.7 cm$^{-1}$ for the i-p and o-p modes, respectively. Moving to higher levels of theory, we obtained 659.6 and 660.3 cm$^{-1}$ using fc-CCSD(T)/ccpVTZ, and 662.3 and 663.8 cm$^{-1}$ using fc-CCSD(T)/aug-cc-pVTZ. All the calculations gave values close to 1.0 for the Coriolis zeta parameter as we expected. In addition, all these calculations agreed with experiment in sign (o-p above i-p) and (approximately) magnitude (values from 0.7 to 1.5 cm$^{-1}$), showing that theory at the harmonic level does capture the essential result. But the calculation at the highest level (CCSD(T)/aug-cc-pVTZ) was actually the worst, showing that there is at least some accidental cancellation of errors going on. The problem, of course, is that we are dealing with a floppy, weakly-bound, van der Waals complex. So even though the intramolecular $CO_2$ bending motion is reasonably harmonic, the intermolecular $CO_2$-Ar stretching and (especially) bending modes are not. A full calculation will first require reliable *ab initio* calculation of the $CO_2$-Ar intermolecular potential as a function of the intermolecular distance and angle, and then a detailed dynamical calculation, essentially integration of the $CO_2$ $v_2$ mode splitting over these two intermolecular motions.

### 3.3. Conclusions

Since we now have parameters for $CO_2$-Ar in the $CO_2$ $(01^10)$ state which are complete except for X, it is possible to predict the $CO_2$-Ar spectrum in the region of the $CO_2$ $v_2$ fundamental band, as shown in Fig. 3. Here, we assume a rotational temperature of 1.8 K and also assume equal transition moments for the i-p and o-p components, which have *a*- and *c*-type selection rules, respectively. As we know the two components are highly mixed, so the spectrum is distorted from a normal appearance and many individual transitions have a contribution from both components. The most prominent features are the *P*- and *R*-branches of the i-p $K = 0 \leftarrow 0$ subband, the *Q*- branch of the o-p $K = 1 \leftarrow 0$ subband, and the *R*-branch of the o-p $K = 3 \leftarrow 2$ subband. In this simulation, we assume that X, the overall vibrational shift of the i-p $v_2$ mode in $CO_2$-Ar, is zero. So the real spectrum, when it is observed, will be moved up or down by X cm$^{-1}$ compared to that in Fig. 3.

The present results show that, under suitable jet expansion conditions, sufficient population remains in the first excited bending state of $CO_2$ $(01^10)$ to enable observation of the $(01^11) \leftarrow (01^10)$ hot band of $CO_2$-Ar, and thus to determine the splitting of the degenerate bending mode into in-plane and out-of-plane components. Similar results should be possible for other weakly-bound complexes containing $CO_2$, and perhaps for complexes containing related linear triatomic molecules such as $N_2O$, OCS, and $CS_2$. This opens up an aspect of intermolecular interactions for which there has been little or no previous theoretical consideration.

**Acknowledgements**

The financial support of the Natural Sciences and Engineering Research Council of Canada is gratefully acknowledged.

Table 1. Molecular parameters for the (011) ← (010) hot band of $CO_2$ - Ar (in $cm^{-1}$).[a]

|  | (000) (Ref. 6) | (010) i-p | (010) o-p | (011) i-p | (011) o-p |
|---|---|---|---|---|---|
| $\sigma_0$ | 0.0 | X [a] | 0.87727(15)+X | 2336.1692(1)+X | 2337.0455(1)+X |
| $A$ | 0.397087 | 0.397769(35) | 0.396917(19) | 0.394624(20) | 0.393911(33) |
| $B$ | 0.066012 | 0.0659192(99) | 0.0660745(92) | 0.065763(14) | 0.0661385(82) |
| $C$ | 0.056143 | 0.0561461(94) | 0.0561461(94) | 0.0561208(53) | 0.056146(10) |
| $10^6 \, D_{JK}$ | 12.0 |  |  | 7.4(13) | 20.5(26) |
| $10^7 \, D_J$ | 5.66 | 4.23(79) | 70(10) | 5.55(46) | 5.25(63) |
| $\xi_b$ |  | 0.133735(13) |  | 0.133561(37) |  |

[a] X is equal to the free $CO_2$ $\nu_2$ frequency (667.380 $cm^{-1}$) plus or minus an unknown vibrational shift which is unlikely to be more than a few $cm^{-1}$.

**Figure Captions**

Fig. 1.  Observed and simulated spectra of $CO_2$-Ar in the region of the $CO_2$ $(01^11) \leftarrow (01^10)$ hot band. Gaps in the observed spectrum correspond to regions of $CO_2$ monomer absorption. Arrows indicate the first line(s) of the *R*-branch (for $\Delta K = +1$ subbands) or *P*-branch (for $\Delta K = -1$ subbands). Division of the spectrum into in-plane and out-of-plane components is somewhat arbitrary because of the large Coriolis mixing of these modes.

Fig. 2.  Rotational energy levels of $CO_2$-Ar in the $CO_2$ $(01^10)$ (left) and $(01^11)$ (right) vibrational states, illustrating the large Coriolis interaction between in-plane and out-of-plane modes. These are "reduced" energies levels with the basic rotational dependence, $\frac{1}{2}(B + C) \times J(J + 1)$, subtracted for clarity (with $\frac{1}{2}(B + C) = 0.061078$ cm$^{-1}$). The in-plane mode is shown in black (Coriolis interaction on) and blue (Coriolis interaction off). The out-of-plane mode is shown in red (Coriolis interaction on) and green (Coriolis interaction off). The energy zeros are the in-plane mode origins from Table 1, equal to X cm$^{-1}$ for the $(01^10)$ state or $(2336.1692 + X)$ cm$^{-1}$ for the $(01^11)$ state. In the lower $(01^10)$ state, the $K = 1$ and 2 levels lie close together and experience particularly large Coriolis mixing. In the upper $(01^10)$ state, the interaction between $K = 0$ and 1, and between $K = 2$ and 3, become more important.

Fig. 3  Predicted spectrum of $CO_2$-Ar in the region of the $CO_2$ $\nu_2$ fundamental band at a temperature of 1.8 K. Here it is assumed that the unknown shift (X in Table 1) of the in-plane mode relative to the free $CO_2$ molecule is zero. So the actual spectrum, which is not yet observed, may be shifted up or down by one or two cm$^{-1}$ from this simulation.

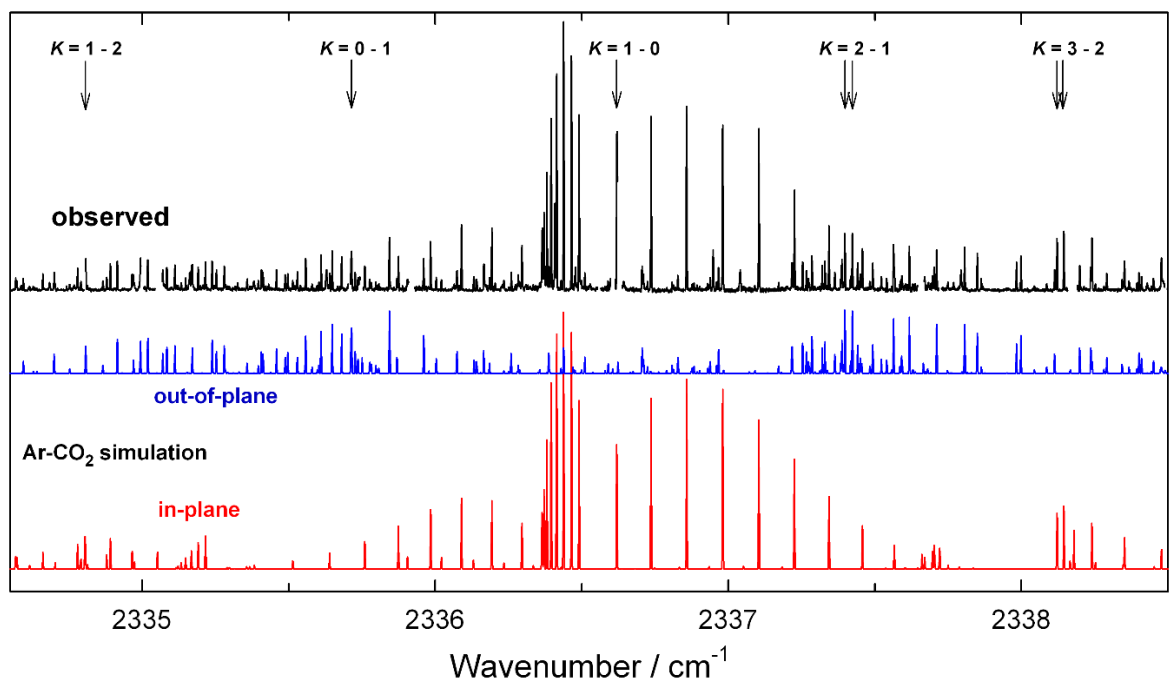

Fig. 1

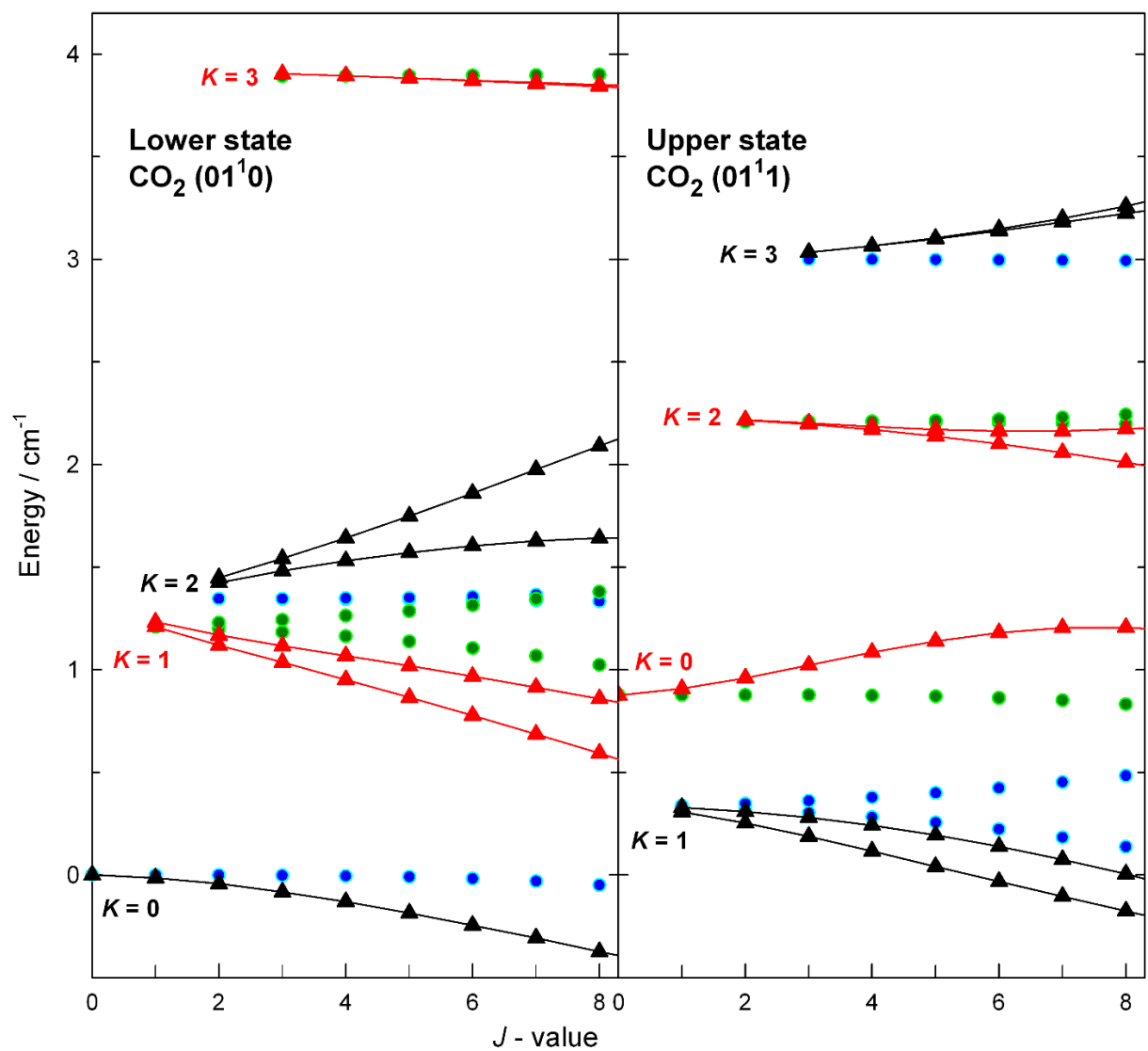

Fig. 2

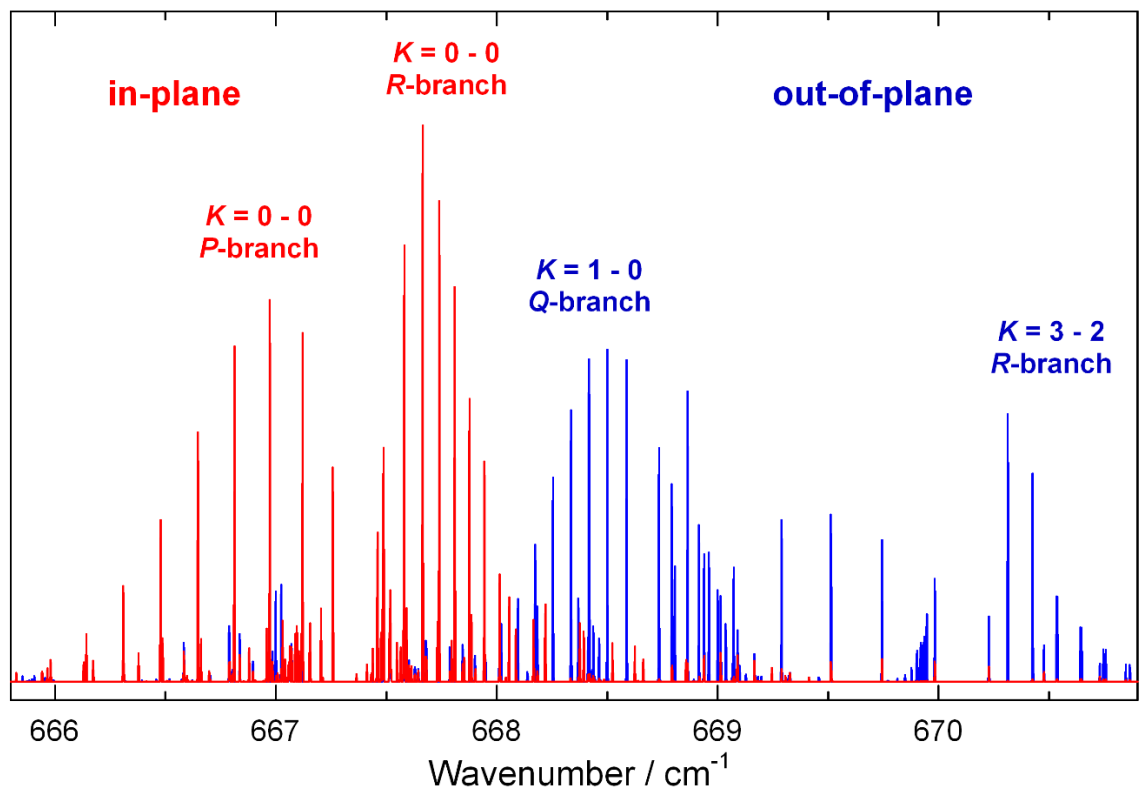

Fig. 3.